\documentclass[aps, prl, twocolumn,amsmath,amssymb,showpacs,groupedaddress]{revtex4-1}
\usepackage{graphicx}

\begin{document}

\preprint{}

\title{Spontaneous symmetry breaking in an optomechanical cavity}

\author{Alexander K. Tagantsev}
\email{alexander.tagantsev@epfl.ch}
\affiliation{Swiss Federal Institute of Technology (EPFL), School of Engineering, Institute of Materials Science, CH-1015 Lausanne, Switzerland}
\affiliation{Ioffe Phys.-Tech. Institute, 26 Politekhnicheskaya, 194021, St.-Petersburg, Russia}

\begin{abstract}
A theoretical consideration of the so-called "membrane-in-the-middle" optomechanical cavity revealed that it undergoes a spontaneous symmetry breaking as a function of transparency of the membrane.
Such typical features of this phenomenon as a square-root development of the order parameter and divergence of the critical susceptibility were identified.
In the contrast to a classical spontaneous-symmetry-breaking system of ferroelectrics, in the system considered, this divergence  remains, due to interference effects, an "internal" property of the system, which does not reveal itself in any singularity of  the output optomechanical response, though the latter is appreciably affected.
\end{abstract}

\date{\today}

\maketitle
Spontaneous symmetry breaking is a process, by which a physical system in a symmetric state ends up in an asymmetric state.
Such an evolution is characterized by the so-called \textit{order parameter}, which is zero in the symmetric state acquiring non-zero values in the asymmetric state.
One also speaks about the appearance of non-zero order parameter as a phase transition.
This is a general  phenomenon, the manifestations of which span from the particle physics~\cite{higgs1964} and cosmology~\cite{bergstrom2006} to the condensed matter physics~\cite{kittel1976}, where ferroelectricity is a classical example.
An important variable used for the description of this phenomenon is the so-called \textit{conjugated field}.
The conjugated field is a perturbation, which induces a non-zero order parameter in the symmetric state.
A characteristic feature of spontaneous symmetry breaking is a divergence (or a strong increase in the case of a discontinuous transition)
of the susceptibility of the order parameter to the conjugated field at the breaking point, which is called \emph{critical susceptibility}.
It is this feature of spontaneous symmetry breaking that is behind most of the applications of ferroelectrics, profiting from highly enhanced dielectric constant, which plays the role of the critical susceptibility.

The \emph{parity-time symmetry breaking in "gain-loss" systems} is currently a hot topic in optics~\cite{ruter2010,kepesidis2016, ozdemir2019}.
In this paper, we theoretically analyze the performance of a simple optomechanical cavity, which exhibits \emph{only dissipation}, to identify a \emph{spontaneous symmetry breaking in the spatial field distribution} in the cavity.
Here the difference of decay rates of two optical modes plays  the role of the order parameter while the mechanical displacement plays the role of the conjugated force.
As a result, the divergence of the critical susceptibility translates into anomalously large dissipative optomechanical coupling constants of the optical modes.
Such a situation, being similar to that in ferroelectrics, however, is yet essentially different: The divergences of the coupling constants of the individual modes do not translate into any singularity of the total optomechanical response, though the latter is appreciably affected.

\begin{figure}
\includegraphics [width=0.7\columnwidth,clip=true, trim=0mm 0mm 0mm 0mm] {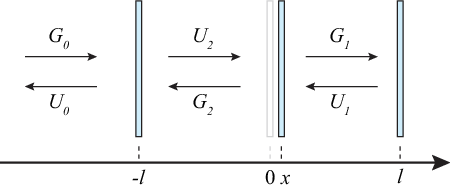}
\caption{Schematic of a membrane-in-the-middle optomechanical cavity.
The left mirror is semitransparent, the right mirror is perfectly reflecting.
The membrane is shown shifted from the middle of the cavity by distance $x$.
Running electromagnetic waves are schematically shown with arrows and labeled  with their complex amplitudes.
\label{f1}}
\end{figure}

The system addressed is the so-called "membrane-in-the-middle" optomechanical cavity, which has been attracting appreciable attention of theorists~\cite{jayich2008,miao2009,genes2013,yanay2016} and experimentalists~\cite{jayich2008,Thompson2008,wilson2009,Purdy2013,mason2019,kampel2017,higginbotham2018}.
It is schematically depicted in Fig.\ref{f1}.
For the case where the membrane is set half-way between the end mirrors, we are interested in the resonance frequencies, decay rates, and constants of optomechanical coupling of the optical modes as well as in the optomechanical signal in the light back-scattered from the cavity.

To be specific, we set the following scattering matrices:
\begin{equation}
\label{mirrorL}
 \left(
  \begin{array}{cc}
 i\tau &  -\rho \\
-\rho  & i\tau \\
  \end{array}
\right),
 \left(
  \begin{array}{cc}
0 &  -1 \\
-1  & 0 \\
  \end{array}
\right), \, \textrm{and}
\left(
  \begin{array}{cc}
 it &  -r \\
-r  & it  \\
  \end{array}
\right)
\end{equation}
for the left mirror, right mirror, and membrane, respectively, where the amplitude transmission coefficients are on the diagonals.
We set $\rho$, $\tau$, $r$,  and $t$ as real and positive ($r^2 +t^2=1$ and $\rho^2 +\tau^2=1$).

Consider the eigen  modes of the system.
Thus, setting  the amplitude of the input field $G_0=0$ (see Fig.\ref{f1}), the complex amplitudes  $G_1$, $G_2$, $U_1$, and $U_2$ (values at the membrane) are linked with the following relations
\begin{align}
\begin{aligned}
\\&G_1=itU_{2} - r U_{1}
\\&G_{2}=-r U_{2}+itU_{1}
\\& G_1e^{ik(l-x)}= - U_{1}e^{-ik(l-x)}
\\& \rho G_2e^{ik(l+x)}= -  U_{2}e^{-ik(l+x)}
\end{aligned}
\label{set}
\end{align}
where for the definition of $l$ and $x$ see Fig.\ref{f1}.
These relations imply the following equation for the resonance $k$-vector
\begin{equation}
\label{condRED}
(e^{-2ikl}-re^{-2ikx})(\rho^{-1}e^{-2ikl}-re^{2ikx})+t^2=0.
\end{equation}
In the absence of dissipation, i.e. at $\rho=1$, Eq. (\ref{condRED}) determines a well-known relation \cite{jayich2008}, $\cos2kl=r\cos2kx$, for a real resonance wavevector.
The complex wave vector $k$, which satisfies (\ref{condRED}), defines the resonance frequencies, decay rates, and optomechanical coupling constants (dispersive and dissipative~\cite{Elste2009,Xuereb2011,Weiss2013}) as follows
\begin{equation}
\label{frequency}
\omega_c =c\textrm{Re}[ k],\qquad\gamma =-2c\textrm{Im}[k],
\end{equation}
\begin{equation}
\label{go}
\frac{d\omega_c}{dx} =c\textrm{Re}\left [\frac{dk}{dx}\right ],\qquad \textrm{and}\qquad \frac{d\gamma}{dx}=-2c\textrm{Im}\left [\frac{dk}{dx}\right ],
\end{equation}
respectively.
Below we study these parameters as functions of $t$ and $\tau$ for the central position of the mirror, i.e. at $x=0$.
We address the case of practical interest where $\tau\ll1$, in what follows, usually keeping only the lowest order terms in this parameter.

At $x=0$, the solutions to (\ref{condRED}) reads
\begin{equation}
\label{solution1}
e^{-2ikl} =  \frac{r(1+\rho)\pm \sqrt{(1-\rho)^2-(1+\rho)^2t^2}}{2}.
\end{equation}
where $\pm$ corresponds to the two modes of the doublets merging in the limit of non-transparent membrane~\cite{footnote10}, i.e. at $t=0$.
A remarkable feature of this solution is that at $t>t_0\equiv(1-\rho)/(1+\rho)\approx\tau^2/4\ll 1$, the square root in (\ref{solution1}) is imaginary, implying the same damping rate for the modes of the doublets while, in the opposite case, it is real, implying, in turn, the degeneracy of the doublet frequencies.
These results are exact.

\begin{figure}
\includegraphics [width=0.7\columnwidth,clip=true, trim=0mm 0mm 0mm 0mm] {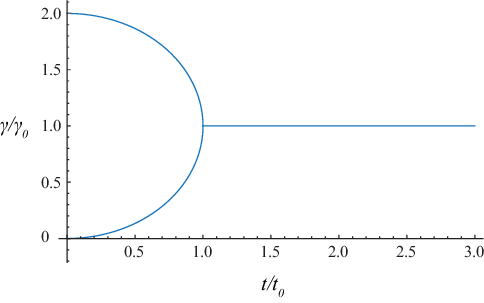}
\caption{Decay rates of two optical modes of the doublet as  functions of the membrane transmission.
Typical square-root development of the order parameter $\gamma-\gamma_0$ is seen.
\label{f2}}
\end{figure}

For the decay rates, using (\ref{solution1}) and (\ref{frequency}),  straightforward calculations yield,
at $t>t_0$,
\begin{equation}
\label{decay11}
\gamma_0=\frac{c\tau^2}{4l}
\end{equation}
while, at $t<t_0$,
\begin{equation}
\label{decay22}
\gamma_{+,-}=\gamma_0[1\mp\sqrt{1-(t/t_0)^2}].
\end{equation}
The $t$-dependence of the cavity decay rates according to (\ref{decay11}) and (\ref{decay22}) is shown in Fig.\ref{f2}.
The spontaneous symmetry breaking behaviour with the typical square-root development, c.f. Ref.~\citenum{strukov2012}, of the order parameter $\gamma-\gamma_0$ in the "asymmetric" state, i.e. at $t<t_0$, is seen here.
Such a behavior is similar to that for the parity-time symmetry breaking in a meta-material-involved system~\cite{ozdemir2019}.
Though the systems are very different, the mathematical descriptions are alike and can be viewed in terms of the so-called exceptional point~\cite{ozdemir2019}.
One should also note that the mathematical framework of the exceptional point was also applied to the description of optomechanical phenomena however in a different physical context:
In Ref.~\citenum{xu2016} a non-reciprocal energy transfer between 2 mechanical modes was addressed while Refs.~\citenum{jing2017} and \citenum{djorwe2019} were dealing with a "gain-loss" situation~\cite{kepesidis2016} , like in the meta-material-involved systems.

The symmetry breaking in our system can also be identified in the field profiles of the modes.
Such profiles  calculated neglecting the energy loss from the input mirror (i.e. at $\tau=0$) are known~\cite{jayich2008,yanay2016} to be  symmetric or antisymmetric with respect to the middle of the cavity.
For non-zero $\tau$, this, however, is not always the case.
We characterize the field profiles with the amplitudes  $G_1$ and $G_2$.
From Eqs.(\ref{set}) one readily finds \cite{footnote9}
\begin{equation}
\label{GG}
\frac{G_2}{G_1}=\frac{r(1-\rho)\pm \sqrt{(1-\rho)^2-(1+\rho)^2t^2}}{2it\rho}.
\end{equation}

In the "symmetric" state, i.e. at $t>t_0$, Eq.(\ref{GG}) yields
$
|G_2/G_1|^2=1/\rho\approx1
$
for both modes of the doublets, reproducing (to within the accuracy accepted) the results of the dissipation-free calculations.
Note that this relation, being  valid for the both modes,  actually means that the energy fluxes in the modes towards the input mirror are the same if the energy stored in the modes are the same also.
This implies the same decay rates of the modes.

In the "asymmetric" state, i.e. at $t<t_0$, Eq.(\ref{GG}) results in an asymmetric field profiles.
Now to within the accepted accuracy of calculations one finds
\begin{equation}
\label{GGratio1}
\left|\frac{G_2}{G_1}\right |^2=\left(\frac{t_0}{t}\right)^2\left[1\mp \sqrt{1-(t/t_0)^2}\right].
\end{equation}
Notably, for the modes corresponding to $\mp$ in this equation, it implies
$G_1^{+}\Rightarrow 0$ and $ G_2^{-}\Rightarrow 0 $ at $t\Rightarrow 0$,
respectively, suggesting that, $t=t_0$, there appears  an onset of localization of the modes into the halves of the cavity.
It is also evident that these relations are qualitatively consistent with (\ref{decay22}).

\begin{figure}
\includegraphics [width=0.7\columnwidth,clip=true, trim=0mm 0mm 0mm 0mm] {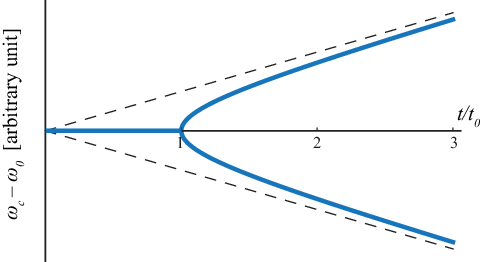}
\caption{Cavity mode splitting vs the membrane transmission for small $t$ . Results of exact calculations - solid lines. Those obtained neglecting the spontaneous symmetry breaking - dashed lines.
\label{f3}}
\end{figure}

The spontaneous symmetry breaking revealed above in the mode decay rates and field profiles also manifests itself in the mode frequencies.
Since the spontaneous symmetry breaking occurs at  $t=t_0 \ll 1$, the further analysis is done in the approximation of small $t$.
Let us consider the modes of a doublet, which at $t\rightarrow 0$ merge into a single mode with frequency  $\omega_0$ (evidently $e^{2i\omega_0l/c}=1)$.
In the absence of dissipation one readily finds~\cite{jayich2008} for the frequencies of these modes
\begin{equation}
\label{split}
 \omega_c -\omega_0=\mp\Omega_0\qquad \Omega_0\equiv\frac{ct}{2l}
\end{equation}
However, the incorporation of the dissipation dramatically affects this result: Eqs.~(\ref{solution1}) and (\ref{frequency}) imply that, at $t<t_0$, as was mentioned above, the doublet stays degenerate at finite $t$, i.e.
$
 \omega_c=\omega_0
$,
while at $t>t_0$,
\begin{equation}
\label{split2}
 \omega_c - \omega_0=\mp\frac{c\sqrt{t^2- t_0^2}}{2l}.
\end{equation}
The cavity mode splitting for small $t$ is schematically illustrated in Fig.\ref{f3}.
Such a behavior is similar to that for the parity-time symmetry breaking in a meta-material-involved system~\cite{ozdemir2019}.

To address the critical anomaly associated with the spontaneous symmetry breaking, the conjugated force is to be identified.
For the order parameter $\gamma-\gamma_0$, a displacement of the membrane from the middle of the cavity $x$ (see Fig.\ref{f1}) can be taken as a conjugated force such that the dissipative coupling constant $\frac{d\gamma}{dx}$ appears to play the role of the critical susceptibility.

\begin{figure}
\includegraphics [width=0.7\columnwidth,clip=true, trim=0mm 0mm 0mm 0mm] {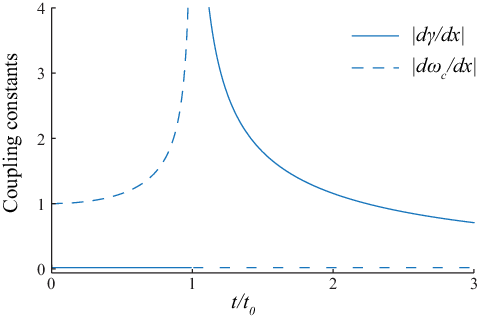}
\caption{Absolute values of optomechanical coupling constants of modes of the doublet normalized to the dispersive coupling constant of the system at $t=0$.
Dispersive coupling - dashed lines.
Dissipative coupling - solid lines.
Nonzero values of the constants where these are relatively small are shown as zero.
\label{f4}}
\end{figure}

Using (\ref{condRED}) and (\ref{solution1}), one finds
\begin{equation}
\label{derX}
\frac{dk}{dx} =\pm\frac{r}{l}\frac{k}{\sqrt{1-t^2\frac{(1+\rho)^2}{(1-\rho)^2}}}.
\end{equation}
Next, to within the accepted accuracy, using (\ref{go}), Eq.(\ref{derX}) implies the following:
at $t>t_0$,
\begin{equation}
\label{derG1}
\frac{d\gamma}{dx}=\pm\frac{2\omega_{0}}{l}\frac{1}{\sqrt{(t/t_0)^2-1}}.
\end{equation}
while, at $t<t_0$,
\begin{equation}
\label{derG2}
\frac{d\gamma}{dx}=\mp\frac{\gamma_{+,-}}{l}\frac{1}{\sqrt{1-(t/t_0)^2}}.
\end{equation}
The dependence of the absolute value of dissipative coupling constants of the doublet modes given by Eqs.(\ref{derG1}) and (\ref{derG2}) is shown in Fig.\ref{f4} where at $t<t_0$ relatively small values of $d\gamma/dx$ are shown as zero.
Here, another characteristic feature of the spontaneous symmetry breaking - divergence of the critical susceptibility - is seen, c.f. Ref.~\citenum{strukov2012}.

One sees that, in the symmetric state, this value may readily exceed the typical value of the dispersive coupling constant for an optomechanical Fabry-Perot cavity of the same length.
Note that even not very close to the transition the absolute values of the dissipative optomechanical coupling constant are much larger than  those discussed in the literature~\cite{Xuereb2011,sankey2010}.

It is instructive to compare the aforementioned result with those from Ref.\cite{genes2013}.
That work deals with the transmission of the system shown in Fig.\ref{f1} where, however, the right mirror is set identical to the left one.
In this system no symmetry breaking happens and the doublet modes are never degenerate, just approaching each other in the limit $t\rightarrow0$ (in our notations) and essentially overlapping at $t<2t_0$.
Here, no critical behavior of optimechanics of individual modes at $t$ close to $t_0$ takes place.
Instead, the paper predicts an anomalously strong non-linear optomechanical response at $t \ll t_0$ (in our notations).

As it is clear from (\ref{derX}), the dispersive coupling constants of the doublet modes also exhibit an anomaly, implying via  (\ref{go}), to within the accepted accuracy,
at $t<t_0$:
\begin{equation}
\label{derD1}
\frac{d\omega_c}{dx}=\pm\frac{\omega_{0}}{l}\frac{1}{\sqrt{1-(t/t_0)^2}}
\end{equation}
while, at $t>t_0$:
\begin{equation}
\label{derD2}
\frac{d\omega_c}{dx}=\mp\frac{\gamma_{+,-}}{2l}\frac{1}{\sqrt{(t/t_0)^2-1}}.
\end{equation}
The dependence of the dispersive optomechanical coupling constants given by Eqs.(\ref{derD1}) and (\ref{derD2}) is schematically shown in Fig.\ref{f4} where, at $t>t_0$, relatively small values of $d\omega_c/dx$ are shown as zero.

In view of the opposite sign of the coupling constants of modes of the doublet, the divergences identified above do not necessarily mean that in the optomechanical signal of the light reflected from the cavity.
To check this, such a signal was calculated using the so-called input-output relations approach~\cite{Buonanno2003,Danilishin2012,Khalili2016}.

Using this approach, we considered the system, schematically shown depicted in Fig.\ref{f1}, to be pumped from the left mirror with a strong monochromatic light of frequency $\omega_L$ and amplitude $G_{00}$.
We are interested in modifications of the light scattered back from the cavity, which are caused by small and slow displacements $x(t)$ of the membrane from its central position, assuming $x(\Omega)k_{L}\ll1$ and
$\Omega\ll \omega_{L}$, where $k_{L}=\omega_{L}/c$ and $x(\Omega)$ is the Fourier transform of $x(t)$ at the frequency $\Omega$.
The Fourier transform at the frequency $\Omega$ of the $x$-modulated part of amplitude of the backscattered light reads (see the Appendix)
\begin{equation}
\label{gm}
u_0^{(x)}(\Omega)=-i\frac{8G_{00}k_Lx(\Omega)}{\tau^2}B(k,k_L)
\end{equation}
\begin{equation}
\label{B}
B(k,k_L) = \tau^4\frac{C(k,k_L)}{2D(k)D(k_L)}
\end{equation}
where $k=(\omega_L+\Omega)/c$, $D(z)=r-e^{-2izl}+ \rho(r- e^{2izl})$, and $C(k,k_L) = \cos[(k+k_L)l] -r \cos[(k-k_L)l]$.
Here $|B(k,k_L)|$ has the meaning of the absolute value of the optomechanical signal normalized the maximal absolute value of that for the considered system with the perfectly reflecting membrane.

It is seen from Eqs.(\ref{gm}) and (\ref{B}) that divergence  of the optomechanical signal at $t\approx t_0$ may occur only if  $D(k)D(k_L)$ tends here to zero.
One readily checks that it is not the case, moreover, even no cusp at $t\approx t_0$ in the $u_0^{(x)}$ vs. $t$ dependence is present.
Thus, in the case of optomechanical cavity, such a characteristic feature of spontaneous symmetry breaking as the divergence of the critical susceptibly remains an "internal" property of the system, which does not reveal itself in any divergence of its observables.
This makes a big contrast with  the manifestation of the spontaneous symmetry breaking in ferroelectrics.

The fact that divergences of the optomechanical coupling constants of individual modes of the doublet disappear from the output signal can be easily rationalized.
Evidently, if the frequencies and dampings of the doublet modes were equal, the coupling constants of the modes that differ only in the sign would result in the total cancellation of their contributions to the  optomechanical signal.
In our system, the modes differ either in frequency or in damping such that the full cancellation does not take place.
Instead, at $t>t_0$ where the dampings of the modes are equal while  the frequencies, according to (\ref{split2}), are split by $\delta \omega_c = c\sqrt{t^2- t_0^2}/l$,
one expects the signal to be proportional to $|\delta \omega_c d\gamma /dx|$.
In view of (\ref{derG1}), being equal to $\omega_0^2\tau^2/(2l^2)$ , this product does not contain any singularity.
On the same lines one can show that, at $t<t_0$, the divergence of the dispersive coupling constant  $d\omega_c/dx$ is washed out from the optomechanical output signal also.
Since the frequency and damping in the doublet modes are available from the above text, it is straightforward to rewrite Eq.~(\ref{gm}) in terms of individual-mode contributions using the Langevin equation framework.
Such an analysis fully supports the qualitative arguments given above.
\begin{figure}
\includegraphics [width=0.9\columnwidth,clip=true, trim=0mm 0mm 0mm 0mm] {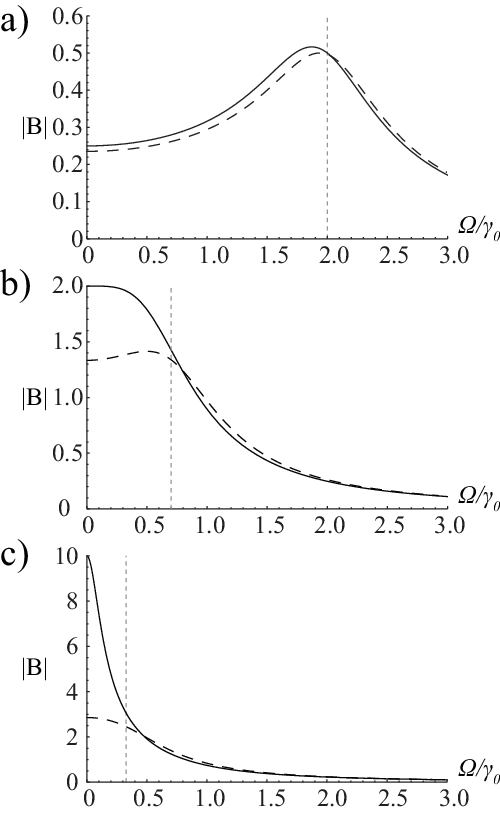}
\caption{Spectrum of absolute value of  optomechanical signal normalized to the maximal absolute value of that for the considered system with the perfectly reflecting membrane (i.e. at $t =0$), denoted as $|B|$:
 (a) - $t=4t_0$, (b)  - $t=1.4t_0$, and (c) - $t=0.65t_0$.
The spontaneous symmetry breaking takes place at $t =t_0$.
Curved dashed lines show the spectrum if the frequency and damping of the modes  were calculated neglecting the spontaneous symmetry breaking.
Vertical dashed lines show the frequencies $\Omega = \Omega_0$  where $2\Omega_0$ is frequency splitting in the doublet calculated neglecting damping, i.e. at $\tau=0$.
The pumping light frequency equals to the resonance frequency at $t=0$, i.e. $\omega_L=\omega_0$.
\label{f5}}
\end{figure}

The absence of singularity associated with the symmetry breaking in the output signal does not mean that there is no manifestation of this phenomenon in it.
In reality, the spectrum of the optomechanical signal is appreciably affected  by the modification of  frequency and damping of the modes of the doublet caused by the symmetry breaking.
As an example we give an expression for $|B(k,k_L)|$ calculated at $k_L=\omega_0/c$  and keeping the lowest terms in $\Omega$, $\tau$, and $t$, which reads (see Appendix)
\begin{equation}
\label{BBB}
|B(\Omega/c+\omega_0/c ,\omega_0/c)| = \frac{\gamma_0^2}{\sqrt{(\Omega^2-\Omega_0^2)^2+\gamma_0^2\Omega^2}}.
\end{equation}
We compare this expression with that where $(\Omega^2-\Omega_0^2)^2+\gamma_0^2\Omega^2$ is replaced with  $|(\Omega-\Omega_0+i\gamma_0/2) (\Omega+\Omega_0+i\gamma_0/2)|^2=(\Omega^2-\Omega_0^2)^2+\gamma_0^2(\Omega^2+\Omega_0^2)/2+\gamma_0^4/16$, the latter corresponding to the frequencies and dampings of the doublet modes calculated neglecting the spontaneous symmetry breaking.
Such a comparison is presented in Fig.\ref{f5}. It is cleanly seen that, far from the summery breaking point, at $t=4t_0$, (a), the optomechanical signal is hardly affected by the symmetry breaking phenomenon while approaching the transition, at $t/\tau^2=1.4t_0$, (b), and further on, at $t/\tau^2=0.65t_0$, (c), the impact is appreciable.

\emph{To summarize.}
A theoretical consideration of the "membrane-in-the-middle" optomechanical cavity revealed that it undergoes a spontaneous symmetry breaking as a function of transparency of the membrane.
Such typical features of this phenomenon as a square-root development of the order parameter and divergence of the critical susceptibility were identified.
In contrast to a classical spontaneous-symmetry-breaking system of ferroelectrics, here this divergence  remains an "internal" property of the system, which does not reveal itself in any divergence of its observables.
At the same time, the spectrum of the optomechanical signal is affected by the phenomenon.

\section*{Appendix: Optomechanical signal in reflected light calculated using the input-output relations approach}
We are interested in an optomechanical signal of the  "membrane-in-the-middle" optomechanical cavity schematically depicted in Fig.\ref{f1} while parameters of the mirrors and membrane are given by Eq.(\ref{mirrorL}).
Specifically, we consider small deviations $x$ of the membrane from its central position and calculate the  $x$-dependent component of the backscattered light when the cavity is exited with a strong coherent light of frequency $\omega_L$.

\subsection{General}
A theory of the system in question was already offered in a number of papers \cite{jayich2008,miao2009,yanay2016}, a comprehensive treatment being given using a perturbation approach \cite{miao2009,yanay2016}.
At the same time, the linear optomechanical problem we are interested in can also be treated practically exactly by using the so-called input-output relations approach \cite{Buonanno2003,Danilishin2012,Khalili2016} popular in the gravitational wave community.
Below we implement such an approach, as yielding a  result, which is free from the limitations of the customary Langevin-equation  formalism.

Following this approach, in the frame rotating with the frequency $\omega_L$,  we present all amplitudes of the fields (see Fig.\ref{f1}) as a sum of a large constant part and a small fluctuating part, e.g.
\begin{equation}
\label{twoparts}
  \begin{array}{cc}
 &  G_{0}(t)=G_{00}+g_0(t) \qquad G_{2}(t)=G_{20}+g_2(t) \\
 &  U_{0}(t)=U_{00}+u_0(t) \qquad U_{2}(t)=U_{20}+u_2(t) \\
  &  \textit{etc}. \\
  \end{array}
\end{equation}
For our system, in view of (\ref{mirrorL}) the following equations
\begin{align}
\begin{aligned}
\\U_{00}=&\,i\tau G_{20}e^{ik_Ll} -\rho G_{00}
\\U_{20}e^{-ik_Ll}=&-\rho G_{20}e^{ik_Ll}+i\tau G_{00}
\\G_{20}=&itU_{10}-rU_{20}
\\G_{10}=&-rU_{10}+itU_{20}
\\G_{10}=&-U_{10}e^{-2ik_Ll}.
\end{aligned}
\label{UG1}
\end{align}
are satisfied for the constant parts, where $k_{L}=\omega_{L}/c$.
The solution to this set of equation reads
\begin{equation}
\label{U201}
U_{10}=-\frac{t\tau e^{ik_Ll}}{D(k_L)}G_{00}
\end{equation}
and
\begin{equation}
\label{U10}
U_{20}=i\frac{\tau e^{ik_Ll}(r-e^{-2ik_Ll})}{D(k_L)}G_{00}
\end{equation}
where
\begin{equation}
\label{D}
D(z)=r-e^{-2izl}+ \rho(r- e^{2izl}).
\end{equation}

The Fourier transforms of fluctuating parts of the amplitudes [denoted as $g_0(\Omega)$ ,$u_0(\Omega)$, etc]. meet the following relations:
\begin{align}
\begin{aligned}
\\u_0(\Omega)=&\,i\tau g_{2}(\Omega)e^{ikl} -\rho g_{0}(\Omega)
\\u_{2}(\Omega)e^{-ikl}=&-\rho g_{2}(\Omega)e^{ikl}+i\tau g_{0}(\Omega)
\\g_{2}(\Omega)=&itu_{1}(\Omega)-ru_{2}+2irU_{20}k_Lx(\Omega)
\\g_{1}(\Omega)=&-ru_{1}(\Omega)+itu_{2}(\Omega)-2irU_{10}k_Lx(\Omega)
\\g_{1}(\Omega)=&-u_{1}(\Omega)e^{-2ikl}.
\end{aligned}
\label{ug1}
\end{align}
where $k=k_L+\Omega/c$ and $x(\Omega)$ is  the Fourier transform of $x(t)$.
Here it is assumed that $x(\Omega)k_{L}\ll1$ and $\Omega\ll \omega_{L}$.

Starting from (\ref{UG1}) and (\ref{ug1}),  the Fourier transform of complex amplitude  of backscattered light, $u_0(\Omega)$, reads
\begin{equation}
\label{u01}
u_0(\Omega)= -\frac{D(k)^*}{D(k)}g_0-iu_m B(k,k_L)
\end{equation}
\begin{equation}
\label{um}
u_m=\frac{8G_{00}k_Lx(\Omega)}{\tau^2}
\end{equation}
\begin{equation}
\label{Bk}
B(k,k_L) = \tau^4\frac{C(k,k_L)}{2D(k)D(k_L)}
\end{equation}
\begin{equation}
\label{C}
C(k,k_L) = \cos[(k+k_L)l] -r \cos[(k-k_L)l].
\end{equation}
Equations (\ref{u01})-(\ref{C}) brings us to Eqs.(\ref{gm}) and (\ref{B}).
\subsection{Nearly resonance excitation}
Consider the case of excitation with a frequency $\omega_L$, which is close to the resonance frequency of the half-cavity $\omega_c$ calculated neglecting the dissipation.
We are infested in  a narrow spectral range about $\omega_L$.
To characterize the detuning with respect to $\omega_c$ and frequency range of interest, which we keep in mind to be about the mechanical frequencies, we introduce the following dimensionless parameters
$$
Q =  (\omega_L-\omega_c)l/c \ll1
$$
and
$$
q =  (k-\omega_L/c)l= \Omega l/c \ll1.
$$
Next we  expand $B(k,k_L)$ with respect to small parameters of the problem $t$, $\tau$, $Q$, and $q$, keeping the lowest order terms.
We readily find
$$
D(k_L)= -t^2+4Q^2+i\tau^2Q,
$$
$$
D(k)= -t^2+4(Q+q)^2+i\tau^2(Q+q),
$$
$$
C(k,k_L)= \frac{t^2-4Q^2-4Qq}{2},
$$
and
\begin{equation}
\label{Bexp}
B(k,k_L) = \frac{\tau^4}{4}\frac{t^2-4Q^2-4Qq}{[t^2-4Q^2-i\tau^2Q][t^2-4(Q+q)^2-i\tau^2(Q+q)]}.
\end{equation}

In the case of the "resonance" excitation, i.e. at $Q=0$, Eq.(\ref{Bexp}) boils down to the from
\begin{equation}
\label{Bexp01}
B(k,k_L) = \frac{\tau^4}{4}\frac{1}{t^2-4q^2-i\tau^2q},
\end{equation}
which, using the definitions $\Omega_0\equiv\frac{ct}{2l}$ and $\gamma_0\equiv\frac{c\tau^2}{4l}$, can be rewritten as
\begin{equation}
\label{Bexp02}
B(k,k_L) = \frac{\gamma_0^2}{\Omega_0^2-\Omega^2-i\gamma_0\Omega_0}.
\end{equation}
This brings us to Eq.(\ref{BBB}).

\begin{acknowledgments}
The author acknowledges fruitful comments and reading the manuscript by Sergey A. Fedorov and  Eugene S. Polzik.
\end{acknowledgments}

\bibliography{QOwork,NF}

\begin{thebibliography}{29}
\bibitem [{foo({\natexlab{a}})}]{footnote10}%
  \BibitemOpen
  \href@noop {} {\bibfield  {journal} {\bibinfo  {journal} {Note that, at
  $t=0$, the system consists of two decoupled half-cavities with coinciding
  resonance frequencies while, at $t=1$, it is just an empty cavity of the
  length $2l$}\ } ({\natexlab{a}})}\BibitemShut {NoStop}%
\bibitem [{foo({\natexlab{b}})}]{footnote9}%
  \BibitemOpen
  \href@noop {} {\bibfield  {journal} {\bibinfo  {journal} {In the absence of
  the dissipation, i.e. at $\rho=1$ , this way we can find the profiles of the
  true physical fields in the system. However, as far as the dissipation is
  involved this is not exactly the case since, in view of complex $k$-vectors,
  our approach is manipulating with the slightly increasing waves between the
  mirrors and membrane instead of perfectly sinusoidal. Nevertheless, such a
  disparity will introduce a relative small mistake about $k_il \cong \tau^2\ll
  1$ such that one can trust the aforementioned calculations to within
  $\tau^2$, which is actually the precision mostly accepted in the work}\ }
  ({\natexlab{b}})}\BibitemShut {NoStop}%
\end{thebibliography}
\end{document}